\documentclass[superscriptaddress,showkeys]{revtex4} 
\usepackage{graphicx,amsmath,latexsym,amssymb}
\usepackage{color}

\usepackage[hidelinks]{hyperref}
\hypersetup{
  colorlinks   = true, 	
  urlcolor     = blue, 	
  linkcolor    = blue, 	
  citecolor   = magenta 	
}

\DeclareMathOperator\erf{erf}

\begin{document}

\title{On Some Quantum Correction to the Coulomb Potential in Generalized Uncertainty Principle Approach}

\author{M. Baradaran}
\email{marzieh.baradaran@uhk.cz}
\affiliation{Department of Physics, Faculty of Science, University of Hradec Kr\'alov\'e,
 Rokitansk\'eho 62, 500 03 Hradec Kr\'alov\'e, Czechia}
\author{L.M. Nieto}
\email{luismiguel.nieto.calzada@uva.es}
\affiliation{Departamento de F\'{\i}sica Te\'{o}rica, At\'{o}mica y \'{O}ptica, and Laboratory for Disruptive Interdisciplinary Science (LaDIS), Universidad de Valladolid, 47011 Valladolid, Spain}
\author{S. Zarrinkamar}
\email{saber.zarrinkamar@uva.es}
\affiliation{Departamento de F\'{\i}sica Te\'{o}rica, At\'{o}mica y \'{O}ptica, and Laboratory for Disruptive Interdisciplinary Science (LaDIS), Universidad de Valladolid, 47011 Valladolid, Spain}
\affiliation{Departament of Physics, Garmsar Branch, Islamic Azad University, Garmsar, Iran}

\date{\today}

\begin{abstract}
Taking into account the importance of the unified theory of quantum mechanics and gravity, and the existence of a minimal length of the order of the Planck scale, we consider a modified Schr\"odinger equation resulting from a generalized uncertainty principle, which finds applications from the realm of quantum information to large-scale physics, with a quantum mechanically corrected gravitational interaction proposed very recently. As the resulting equation cannot be solved by common exact approaches, we propose a Bethe ansatz approach, which will be applied and whose results we will discuss,
commenting on the analogy of the present study with some other interesting physical problems.
\end{abstract}

\keywords{minimal length, generalized uncertainty principle, Planck scale, Schr\"odinger equation, quantum correction, Coulomb potential}

\maketitle

PACS: 04.60.-m, 03.65.Ge.

\section{Introduction}
\label{sect:Intro}

There have been several theories to provide a unified description of quantum mechanics and gravity, including quantum gravity, black hole physics, double special relativity, and string theory, among others (see \cite{Konishi,Maggiore,Cortes,Hossenfelder,Scardigli} and references quoted therein), with each approach having its own strengths and failures. However, the common point in such theories is that they all predict the existence of the so-called minimal length which is on the Planck scale $\ell_p=\sqrt{\hbar G/c^3}$. This minimal length is equivalent to a generalization of the uncertainty principle (GUP) that affects the entire physical system.

From the mathematical point of view, in a GUP formalism the equations become more complicated in structure and finding analytical solutions becomes a really challenging task. In particular, what in these circumstances we could call the Schrödinger equation appears in higher order forms: four, six or more, depending on the choice of the GUP and the operators involved. Obviously, the latter is not a very well studied category in mathematical physics, where in general the differential equations that appear, whether relativistic or non-relativistic, are of the first or second order.

As a consequence, considerable efforts have been made to investigate various aspects of the GUP formalism. For example, Ali et al. \cite{Ali} proposed a GUP compatible with string theory, black hole physics, and doubly special relativity, and calculated several magnitudes, including the Lamb shift and the tunnel current.
A detailed study of the Schr\"odinger equation, with the harmonic oscillator problem as a toy model, was carried out in the work of Kempf and his collaborators \cite{Kempf}. The one-dimensional box and the free particle problems within the GUP approach are studied by Nozari and Azizi \cite{Nozari} by considering an approximation to the associated Schr\"odinger equation.
The exact analytical solutions of the hydrogen atom were obtained by Brau \cite{Brau} and the analysis of the inverse square term was performed by Bouaziz and Bawin in \cite{Bouaziz}.
The dispersion states of the Woods-Saxon potential were investigated in \cite{Maghsoodi}. In the relativistic realm, Hassanabadi et al. \cite{Rajabi} proposed an approximate scheme to study the Dirac equation in the GUP formalism based on the resolution of a Schr\"odinger-like counterpart. Furthermore, the concept of GUP has been considered in the study of the thermodynamics of black holes in very recent articles \cite{Bargueno,Buoninfante,Lambiase}.
It is worth noting that Bishop and his collaborators, in a series of articles, have commented on a very deep conceptual point within the GUP formalism: in \cite{Singleton} it is shown that, contrary to the common thinking, different pairs of modified operators can lead to the same modified commutator and still give a different minimum length or even none at all, and \cite{Bishop} explains exhaustively that the results depend on how the modified operators are defined and that the resulting modified commutator is not the only important aspect.

On the other hand, the Coulomb potential is obviously the most physical interaction in common with gravity and quantum mechanics. 
From a pedagogical point of view, the Coulomb potential is studied very carefully in comparison with other examples of quantum mechanics such a the harmonic oscillator or potentials with exponential-type terms.
This is due to several points, including the singularity of the potential, the strange form of the function, and the behavior of the wave function, which extends over the whole space and affects, for example, the partition function. 
Due to such considerations, it should be emphasized that, contrary to common belief, the quantum Coulomb interaction, even in the apparently simple one-dimensional case, remains an attractive and challenging field of study and very recent articles continue to discuss related properties, more than a hundred years after it was first studied.
It has been shown, through various approaches \cite{Kirilina,Netto,Verch}, that the quantum correction of this potential will contain inverse quadratic, inverse cubic and inverse quartic terms. This detail is of particular interest to us, since what we are going to do in this work is to analyze this type of corrected Coulomb potential that includes negative powers up to the fourth order, which we will call  Coulomb--4 potential, within the GUP formalism.

This work is organized as follows. The Schrödinger equation modified with the GUP formalism is reviewed compactly in Section~\ref{sect:model}. Section~\ref{sect:QESord} investigates the solutions to the ordinary case (without GUP) using the Lie algebraic approach and Heun functions. 
In Section~\ref{sect:BAEGUP} the general solution to the modified problem due to the GUP is first explored and then both the ground state and the first excited state are explicitly determined.
Finally, in the last section the conclusions are presented and possible additional studies along these lines are proposed.

\section{The modified Schr\"odinger equation with a GUP and a quantum correction to the Coulomb interaction}
\label{sect:model}

We consider a GUP of the form \cite{Kempf} 
\begin{equation}
\left[x_{G},p_{G}\right]=i\hbar \left( 1+ \frac{\beta}{2\hbar^2} \ p^2 \right) \,,
\end{equation}
where the original minimal length parameter $0\le\beta\le1$ comes from the associated GUP, a coefficient $\frac{\beta}{2\hbar^2}$ is considered in our calculations instead of the original $\beta$ for further convenience and the generalized $ x$ operator is defined as
\begin{equation}
x_{G}=x,  
\end{equation}
with $x$ and $p$ being the ordinary position and momentum operators, respectively. In one spatial dimension, and neglecting some high order terms, the above GUP corresponds to the modified Schr\"odinger equation \cite{Maghsoodi}
\begin{equation}\label{GUPeq}
\left(-\frac{d^2}{dx^2}+V_{e}(x)\right)\psi_n^{G}(x)=0,
\end{equation}
with an effective potential
\begin{equation}\label{VeffDef}
	V_e(x):= \frac{2m}{\hbar^2} \left(V(x)- E_n^{G}\right) +\left(\frac{2m}{\hbar^2}\right)^2 \beta \left(V(x)- E_n \right)^2,
\end{equation}
in which $E_n$ denotes the ordinary case eigenvalues of the energy (with $\beta=0$) and $E_n^{G}$ the eigenenergies for $\beta\neq 0$. Although it is not explicitly indicated, obviously $E_n^{G}(\beta)$ are some functions of $\beta$ such that $E_n^{G}(0)= E_n$.
Note that, Eq. \eqref{GUPeq} is an approximation to the complete differential equation which appears, which is of order six \cite{Rajabi,Nozari}.

Now we are interested in analyzing a specific problem that could be of great physical interest. 
Based on Refs. \cite{Lambiase,Donoghue,Kirilina,Netto,Verch}, let us now consider the following quantum correction to the Coulomb interaction that includes negative powers up to fourth order, that we will call Coulomb--4 potential:
\begin{equation}\label{potV}
	 \frac{2m}{\hbar^2} V(x)=
	 \frac{\alpha _1}{x}
	 +\frac{\alpha _2}{x^2}
	 +\frac{\alpha _3}{x^3}
	 +\frac{\alpha_4^2}{x^4}
	 ,\quad \alpha_4>0, \  \alpha_1<0,
\end{equation}
which corresponds to the effective potential
\begin{equation}\label{Veff8}
V_{e}(x)=
	 \gamma _0+\frac{\gamma _1}{x}+\frac{\gamma _2}{x^2}+\frac{\gamma _3}{x^3}+\frac{\gamma _4}{x^4}+\frac{\gamma _5}{x^5}+\frac{\gamma _6}{x^6}+\frac{\gamma _7}{x^7}+
\frac{\gamma _8}{x^8}
\,, 
\end{equation}
in which the $\gamma _i$, $i=0,1,\dots 8$ are as follows
\begin{subequations}\label{gammas}
	\begin{align} 
    \gamma _0&= \beta  \epsilon_n^{\,2} - \epsilon_n^{G}  ,       \\
	\gamma _1&=\alpha_1 \big(1-2 \beta  \epsilon_n \big),        \\
	\gamma _2&=\alpha_2+ \beta  \big( \alpha_1^2-2 \alpha_2 \epsilon_n  \big),        \\
	\gamma _3&=\alpha_3+2 \beta \left( \alpha_1 \alpha_2 - \alpha_3   \epsilon_n \right) ,     \\
	\gamma _4&= \alpha_4^2 + \beta \big(2 \alpha_1 \alpha_3+\alpha_2^2-2 \alpha_4^2 \epsilon_n \big), \\
	\gamma _5&= 2 \beta  (\alpha_1 \alpha_4^2+\alpha_2 \alpha_3) ,        \\
	\gamma _6&= \beta   \left(2 \alpha_2 \alpha_4^2+\alpha_3^2\right) ,        \\
	\gamma _7&=2 \beta \alpha_3 \alpha_4^2        ,          \\
	\gamma _8&=\beta \alpha_4^4     ,     
	\end{align}
\end{subequations}
where we have introduced the notation
\begin{equation}
\epsilon_n=  \frac{2m}{\hbar^2}\, E_n ,
\qquad 
\epsilon_n^{G}=  \frac{2m}{\hbar^2} \,
E_n^{G}.
\end{equation}
It should be noted that the problem considered here only makes sense in the half-line,
that is, $x$ is always positive and at the origin it is assumed that there is an impenetrable infinite wall that prevents the passage of the particle in the other direction.
Before continuing, it is worth briefly commenting  on the choice of interaction given in \eqref{potV}. 
The first term is the ordinary Coulomb or gravitational potential. 
The second term is usually considered to be the relativistic or post-Newtonian correction to the gravitational problem and includes the speed of light, $c$.
Regarding the quantum correction terms, there is not a quite unified approach, but almost all existing ones include the inverse cubic term as the necessary quantum correction to the gravitational potential. 
However, as already mentioned, the main quantum effect is included in the term $\alpha _3$ and the correction parameters do not have unique values in the existing literature, which requires further investigations. 
It should be noted that both the sign and the value of the parameters are quite different, and even opposite, in several articles \cite{Hamb95,Don94,Net22}, which motivates further studies. However, in this work we will report what can be derived analytically in each of the  cases that we have analyzed.

Since solutions of the model with GUP depend on the solutions of the associated ordinary case ($\beta=0$), let us first start with the analysis of this one.

\section{Schr\"odinger equation for a Coulomb--4 potential}
\label{sect:QESord}
The ordinary case has already been studied via the ansatz method in \cite{Dong} and also in \cite{AgZh2013,PaBar16} with some additional terms in the potential. Here, we will obtain the general solutions of the model using the Lie-algebraic method within the framework of quasi-exact solvability. Moreover, we will show that under appropriate transformation, the corresponding differential equation can be expressed in the form of the double-confluent Heun (DCH) equation. The Schr\"odinger equation with potential \eqref{potV} appears in the form
\begin{equation}\label{OrdScheq}
	\left(- \frac{d^2}{dx^2}+\frac{\alpha _1}{x}+\frac{\alpha _2}{x^2}+\frac{\alpha _3}{x^3}+
	\frac{\alpha _4^2}{x^4}
	-\epsilon_n \right)\psi_n (x)=0.
\end{equation}
By inspecting the asymptotic behavior of the wave function $\psi_n(x)$, it follows that it is convenient to use the following ansatz \cite{Dong,AgZh2013,Turbiner,Turbiner88}
\begin{equation}\label{WavAnsOrd}
	\psi_n (x)=x^{ \delta}\, \exp\left[ - \left(x\sqrt{-\epsilon_n}+\frac{\alpha _4}{x }  \right)\right]\, \varphi_n(x), 
	\qquad 
	\delta=1+\frac{\alpha _3}{2 \alpha _4}>0 ,
\end{equation}
which transforms Schr\"odinger equation \eqref{OrdScheq} into the form 
\begin{equation}\label{OrdScheqTrans}
 \left\lbrace-x^2\frac{d^2}{dx^2}- {2} \left(  \alpha _4 + \delta\, x- \sqrt{-\epsilon_n}\, x^2\right) \frac{d}{dx}+\left(\lambda_1 \,x+\lambda_2 \right)\right\rbrace\varphi_n (x)=0,
\end{equation}
where
\begin{equation}\label{lambdas}
		\lambda_1= \alpha _1 +\left(2+ \frac{\alpha _3 }{\alpha_4}  \right) \sqrt{-\epsilon_n} ,
		\qquad
	\lambda_2= \alpha _2-\frac{\alpha _3^2}{4\alpha _4^2}-  \frac{\alpha _3}{2\alpha _4} +2\alpha _4  \sqrt{-\epsilon_n}.
\end{equation}


Below we will mention some important details about the analytical solutions found and the restrictions imposed, which will be important for subsequent studies on this model.
For this purpose, we focus on the ordinary case of the ground and the first excited state,  mainly due to the similarity with other known problems that appear as special cases of the problem, although similar arguments are valid for the higher order and the GUP cases.

Let us return to our ordinary case, in which we have considered a one-dimensional generalization of the Coulomb interaction with four terms based on the arguments already mentioned above.
As far as the pure Coulomb term is concerned, there is not much challenge because the solutions, in both one, two and three spatial dimensions, have already been reported by a variety of techniques including factorization, supersymmetric quantum mechanics, series solutions, etc.

However, the problem becomes more complicated in our case, where quantum effects, i.e., inverse cubic and inverse quartic terms, are present.
 In this case, although solutions can be obtained by a couple of non-perturbative techniques, including the Heun function and the Lie algebraic approach  for up to a quantum number,  we have strict restrictions on the potential parameters. 
Such restrictions are not very strange since they are already present in other approaches, including series solutions, but the case of generalized four-term potential considered here is more complicated, and becomes even more cumbersome in the GUP case, where eight terms are present in the effective potential.

Next we are going to analyze the problem from two different and complementary points of view: first the Lie algebraic approach and then we will use the Heun functions.

\subsection{The Lie-algebraic approach}
\label{sect:LieOrd}
Following the standard idea of quasi-exact solvability \cite{PaBar16,Turbiner,Turbiner88,Artemio94,BarPa18}, we find that if the constraint
\begin{equation}\label{QESenercons}
\lambda_1=- 2n \sqrt{-\epsilon_n}
\end{equation}
holds, the equation \eqref{OrdScheqTrans} can then be expressed as a quasi-exactly solvable (QES) differential operator in the Lie-algebraic form \cite{AgZh2013,Turbiner,Turbiner88}
\begin{equation}\label{OrdSchLie}
H_{qes}\;\varphi_n (x)=0,
\end{equation}
with 
\begin{equation}
 H_{qes}=-\mathcal{J}_n^+ \mathcal{J}_n^- +2 \sqrt{-\epsilon_n }\,\mathcal{J}_n^+  -2 \alpha _4\,\mathcal{J}_n^- -\Big(2\delta+n\Big)\mathcal{J}_n^0-\frac {n^2}2-n\delta+\lambda_2\,.
\end{equation}
Here,
\begin{equation}\label{GenerJJJ}
	  \mathcal{J}_n^+ =x^2\,\frac{d}{dx}-n\,x ,\qquad
	  \mathcal{J}_n^0 = x\,\frac{d}{dx}-\frac n2,\ \qquad
      \mathcal{J}_n^- = \frac{d}{dx},
\end{equation}
are the generators of the $sl(2)$ Lie algebra satisfying the commutation relations \cite{PaBar16,Turbiner,Artemio94,Turbiner88,Ronveaux}
\begin{equation}
[\mathcal{J}_n^+,\mathcal{J}_n^-]=-2\mathcal{J}_n^0, \qquad  [\mathcal{J}_n^{\pm},\mathcal{J}_n^0]=\mp\mathcal{J}_n^{\pm}. 
\end{equation}
The operators leave invariant the $(n+1)$-dimensional linear space of polynomials spanned by $\langle 1,x,x^2,\dots,x^n\rangle$. In other words, 
\begin{equation}
\varphi_n (x)=\sum_{k=0}^{n} c_k \,x^k,
\end{equation}
where the coefficients $c_k$ satisfy the three-term recursion relation 
 \begin{equation}\label{recur}
c_{k+1}=\frac{ \left( \lambda_2-2 k\,\delta-k(k-1)\right)c_{k}-4\sqrt{-\epsilon_n  }\,c_{k-1}}{	2  (k+1) \,\alpha _4},
\qquad k=0,1,\dots, n,
 \end{equation}
assuming $c_{-1}=0$ and  $c_{n+1}=0$. Equivalently, the recursion relation \eqref{recur} can be rewritten as a tridiagonal matrix equation the nontrivial solutions of which are determined by
 \begin{equation}\label{matrix}
\left|
\begin{array}{ccccc}
	\lambda_2 & -2\alpha_4 &   &   &   \\ [1ex]
	-2n\sqrt{-\epsilon_n} & (\lambda_2-2\delta) & -4\alpha_4 &   &   \\  [1ex]
	  & -2(n-1)\sqrt{-\epsilon_n} &\ddots & \ddots &   \\ [1ex]
	  &   & \ddots & \ddots & -2 n\,\alpha_4 \\ [1.5ex]
	  &   &   & -2\sqrt{-\epsilon_n} & (n-n^2+\lambda_2-2n\,\delta)
\end{array}
\right| =0.
\end{equation}
Note that the condition \eqref{matrix} imposes severe restrictions on the potential parameters $\alpha_i\,$, $i=1,2,3,4$. 
On the other hand, from \eqref{lambdas}--\eqref{QESenercons},  the following expression of the energy in closed form can be obtained
\begin{equation}\label{QESenerExp}
\epsilon_n = -\frac{\alpha _1^2\,  \alpha _4^2}{(\alpha _3+2 (n+1)\alpha _4)^2},
\end{equation}
provided $\alpha_1<0$, which is concluded immediately after replacing $\lambda_1$ in \eqref{QESenercons} and taking into account that, as we have seen in \eqref{WavAnsOrd}, $\delta=1+\frac{\alpha _3}{2 \alpha _4}>0$. 
Therefore, for any given $n$, the general solutions of the energies and the corresponding wave functions, for the ordinary case, are given respectively by \eqref{QESenerExp} and \eqref{WavAnsOrd} together with \eqref{recur}, as long as the restriction \eqref{matrix} is maintained on the parameters of the potential. Next we will find the explicit solutions of the ground state and the first excited state.

\subsubsection*{\textit{(i)} Ground state}
\label{sect:LieOrdn0}

From \eqref{QESenerExp} and \eqref{WavAnsOrd}, the ground state energy and the associated wave function are  given, respectively, by
\begin{equation}
\label{QESenerExpn0}
\epsilon_0 = -\frac{\alpha _1^2\,  \alpha _4^2}{(\alpha _3+2 \alpha _4)^2},
\end{equation}
and
\begin{equation}
\label{WavAnsOrdn0}
\psi_0 (x)=c_0\ x^{1+ \alpha _3/(2 \alpha _4)}\ \exp\left[ - \left(x\sqrt{-\epsilon_0}+\frac{\alpha _4}{x }  \right)\right] ,
\end{equation}
where, due to \eqref{matrix}, the restriction on the parameters of the potential is determined by $\lambda_2=0$ in \eqref{lambdas} or, more explicitly, by 
\begin{equation}
\label{PotConsOrdn0}
4 \alpha_2\, \alpha_4^2-\alpha_3^2-2 \alpha_3\,\alpha_4+8 \alpha_4^3 \sqrt{-\epsilon_0}=0.
\end{equation}

\begin{figure}[htb]
	\centering
	\includegraphics[width=0.4\textwidth]{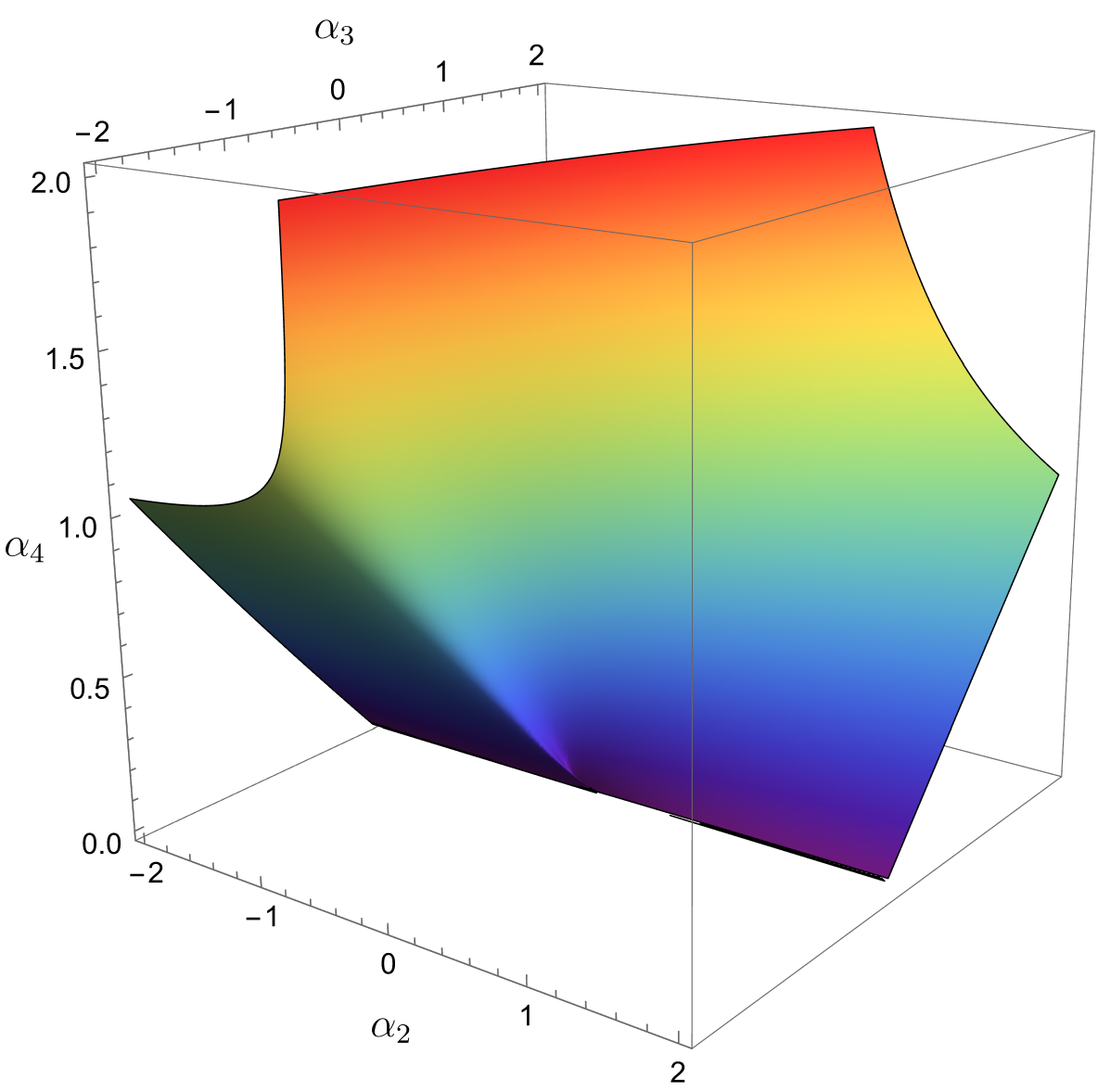}
	\caption{   Allowed values of the potential parameters \eqref{PotConsOrdn0}, corresponding to the ground state \eqref{QESenerExpn0}, for $\alpha_1=-1/{10}$.}
	\label{figparametersground}
\end{figure}

\begin{figure}[htb]
\centering
\includegraphics[width=0.47\textwidth]{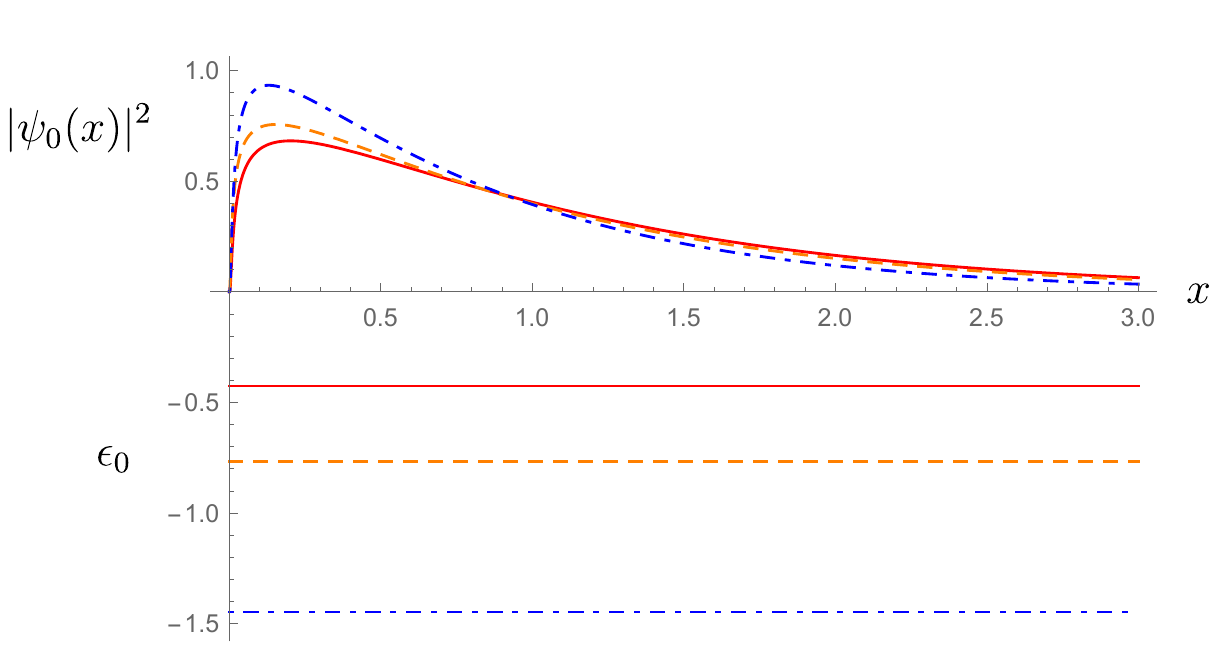} 
\hfill 	
\includegraphics[width=0.47\textwidth]{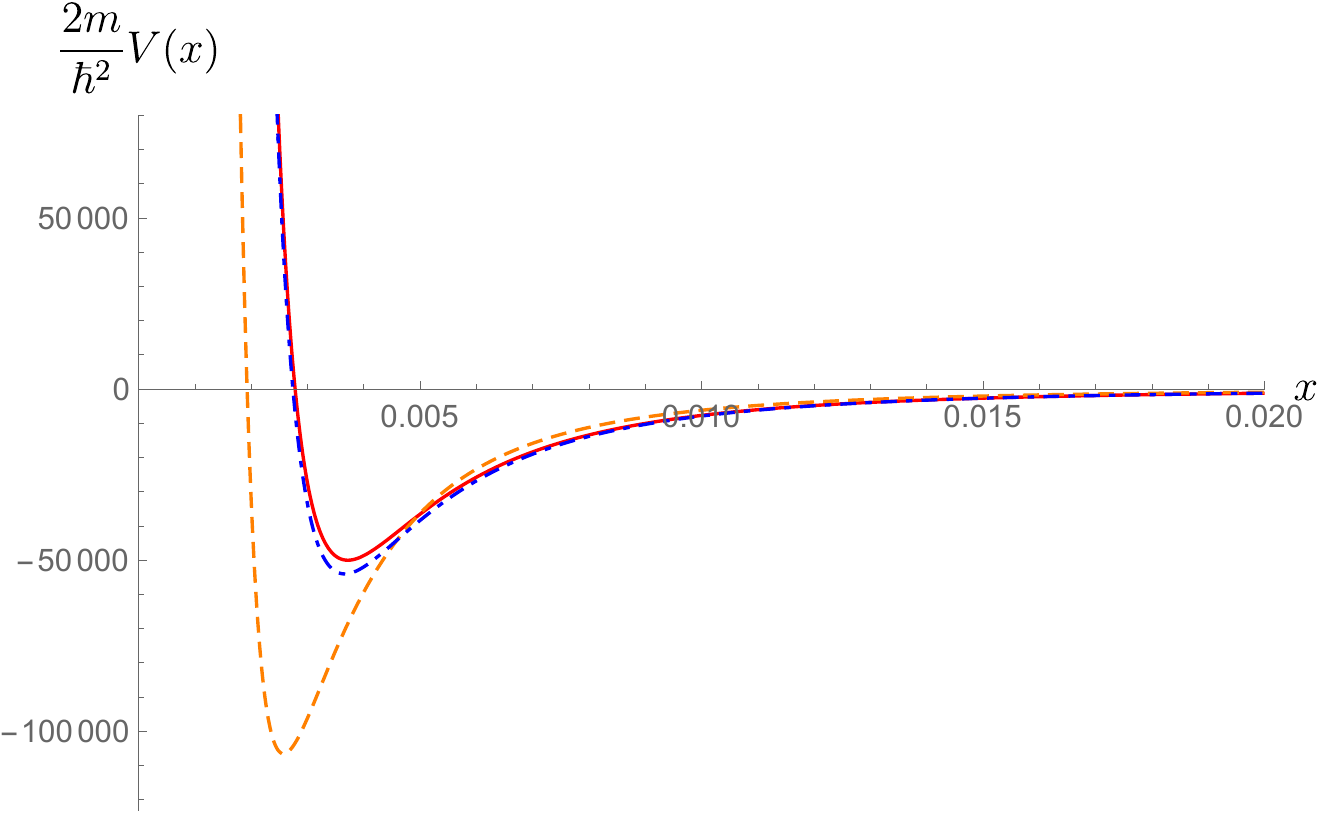}
	\caption{ 
	In the drawing on the left, three values of the negative energy of the ground state and their corresponding probability densities $|\psi_0(x)|^2$, differentiating each case by the color of the lines. Using the same criterion, the corresponding Coulomb--4 potentials~\eqref{potV} are represented in the drawing on the right.
The values of the potential parameters are the following: $\alpha_1=-1/{10}$ in all cases and  
	$(\alpha_2=-0.0776,\alpha_3=-0.0097,\alpha_4=0.0053)$ for the red curves,  
	$(\alpha_2=-0.0603,\alpha_3=-0.0070,\alpha_4=0.0037)$ for the orange curves and  
	$(\alpha_2=-0.0527,\alpha_3=-0.0102,\alpha_4=0.0053)$ for the blue curves,  
	all of them determined from the potential constraint \eqref{PotConsOrdn0}, cf.~FIG.~\ref{figparametersground}. }
	\label{figpotentialandpsi_zero}
\end{figure}

Figure~\ref{figparametersground} shows a graph of the allowed values of the parameters $\alpha_2$, $\alpha_3$ and $\alpha_4$, which are related by \eqref{PotConsOrdn0}, with the restrictions given by the inequalities of \eqref{potV} and \eqref{WavAnsOrd}, which give rise to a fundamental energy state given by the wave function \eqref{WavAnsOrdn0}, with energy \eqref{QESenerExpn0}.
To better capture the meaning of the analytical results found above, the Figure~\ref{figpotentialandpsi_zero} represents, on the one hand, the wave function of the ground state together with the corresponding energy (on the left), and on the other hand, the form of the potential (on the right), all for various values of the parameters $(\alpha_2,\alpha_3,\alpha_4)$.
It should be noted once again that the parameters have been chosen simply to show the limitations of the analytical approach in this case and that normally the main quantum correction coefficient, i.e. $\alpha_3$, is taken as a negative parameter.

\subsubsection*{\textit{(ii)} First excited state}
\label{sect:LieOrdn1}
\begin{figure}[htb]
	\centering
	\includegraphics[width=0.4\textwidth]{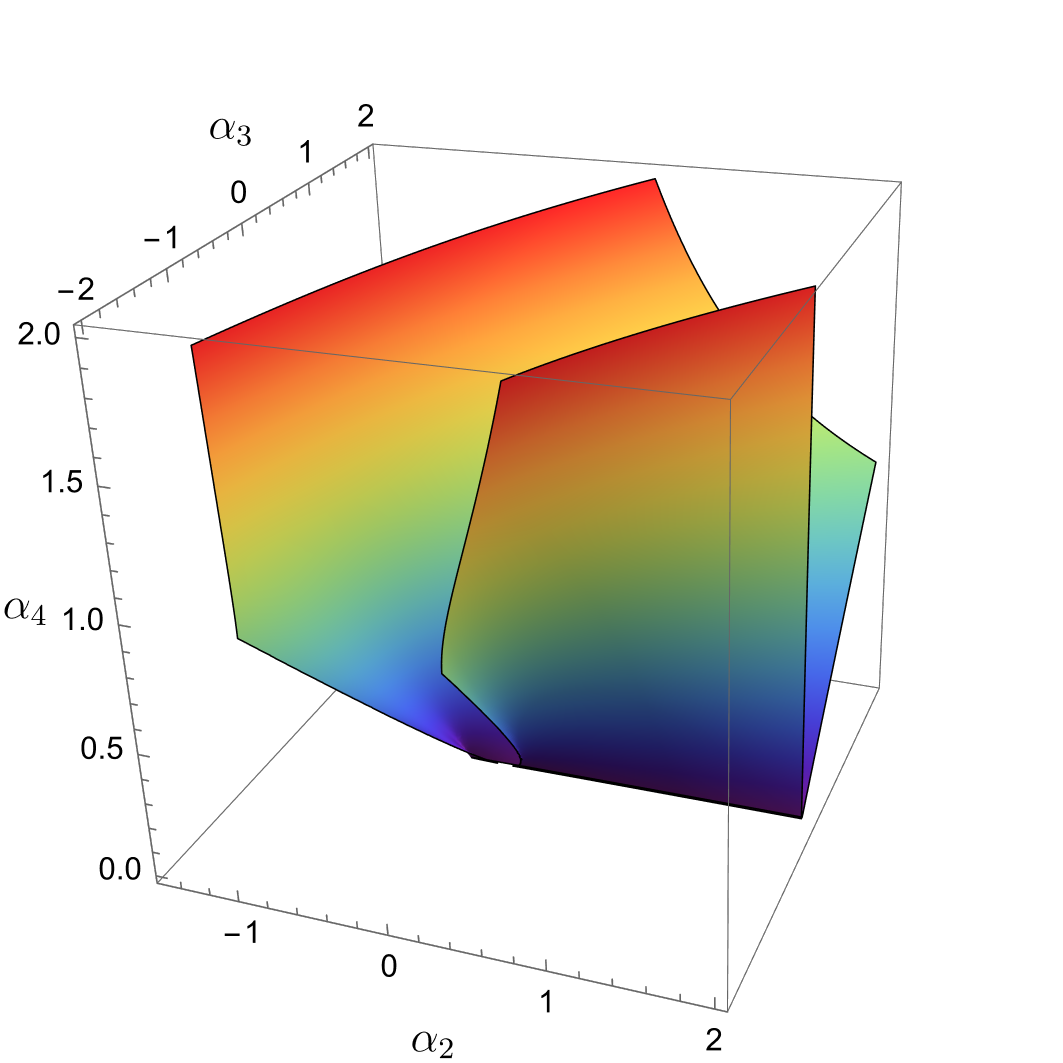}
	\caption{ Allowed potential parameters in \eqref{PotConsOrdn1}, corresponding to the first excited state of the ordinary case, for $\alpha_1=-\frac15$.}
	\label{fig4restrictionsexcited}
\end{figure}

\begin{figure}[htb]
	\centering
\includegraphics[width=0.47\textwidth]{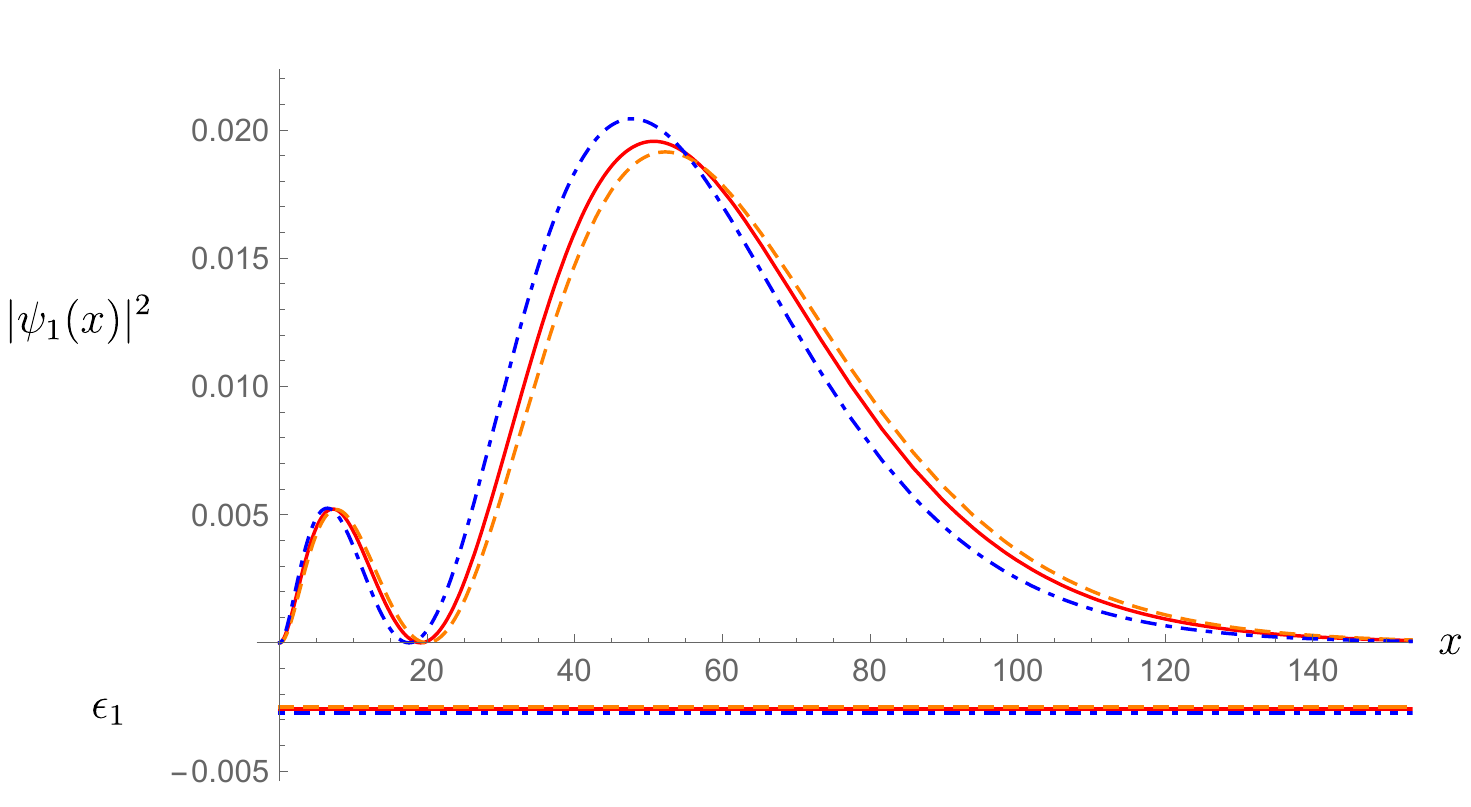}
\hfill
\includegraphics[width=0.47\textwidth]{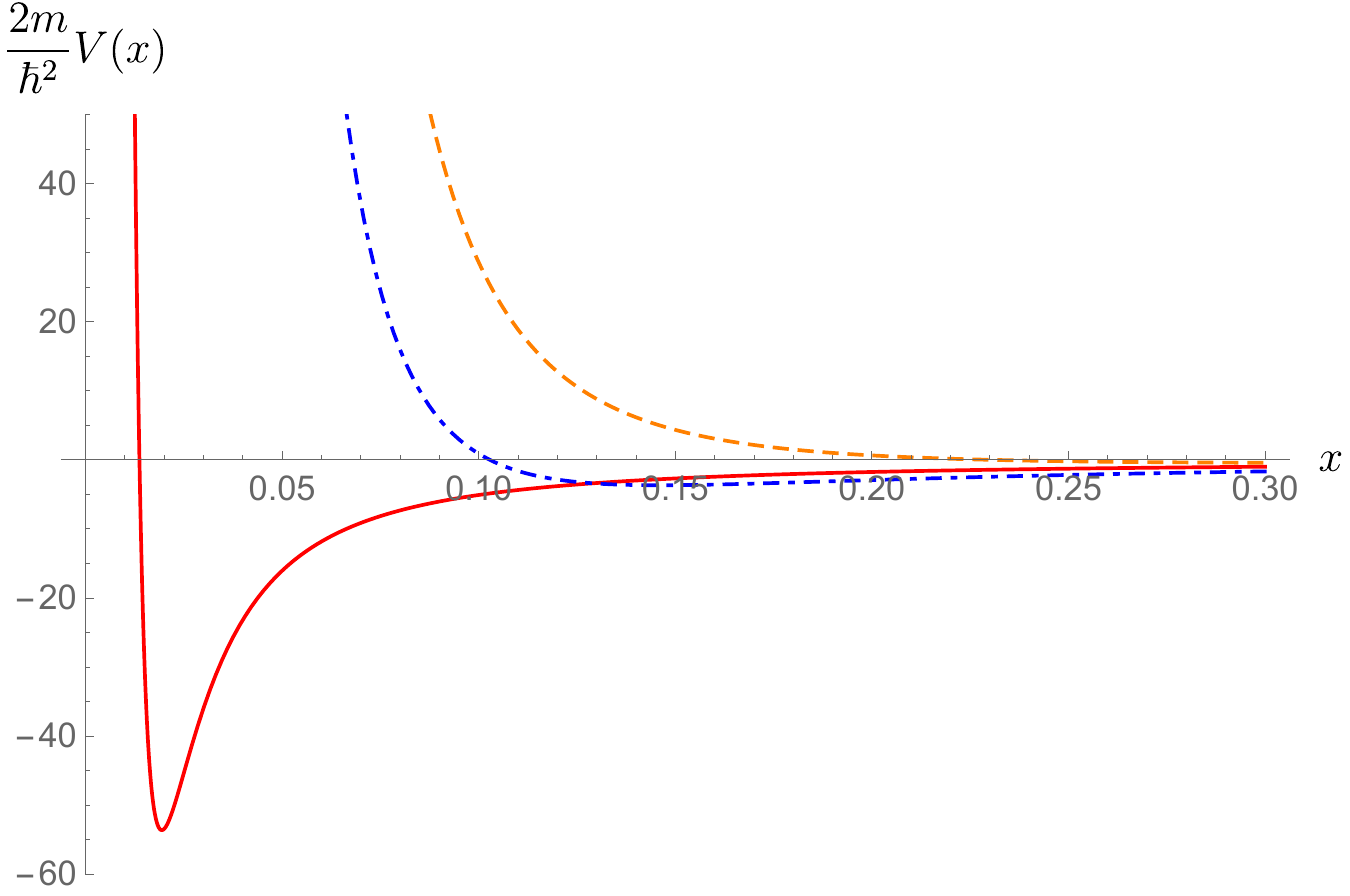} 
	\caption{   On the left, three values of the negative energy of the first excited state and their corresponding probability densities $|\psi_1(x)|^2$, differentiating each case by the color of the lines. Using the same criterion, the corresponding Coulomb--4 potentials~\eqref{potV} are represented in the drawing on the right.
The values of the potential parameters are the following: $\alpha_1=-\frac15$ in all cases and  
	$(\alpha_2=-0.0301,\alpha_3=-0.0002,\alpha_4=0.0029)$ for the red curves,  
	$(\alpha_2=-0.0141,\alpha_3=-0.0003,\alpha_4=0.0569)$ for the orange curves and  
	$(\alpha_2=-0.0880,\alpha_3=-0.0075,\alpha_4=0.0438)$ for the blue curves,  
	all of them determined from the potential constraint \eqref{PotConsOrdn1}, cf.~FIG.~\ref{fig4restrictionsexcited}. }
	\label{figpotentialandpsi_first}
\end{figure}

For $n=1$, from \eqref{QESenerExp} and \eqref{WavAnsOrd}, the energy and the corresponding wave function are  given by
\begin{equation}\label{QESenerExpn1}
	\epsilon_1= -\frac{\alpha _1^2\, \alpha _4^2}{\left(\alpha _3+4 \alpha _4 \right)^2}\,,
\end{equation}
and
\begin{equation}\label{WavAnsOrdn1}
	\psi_1(x)=(c_0+c_1 x)\ x^{1+ \alpha _3/(2 \alpha _4)}\ \exp\left[ - \left(x\sqrt{-\epsilon_1}+\frac{\alpha _4}{x }  \right)\right],
	\end{equation}
in which, from \eqref{recur}, we have
\begin{equation} 
	c_1=\frac{\lambda_2}{2\alpha_4}\, c_0\,.
	\end{equation}
In this case, from \eqref{matrix}, the restriction on the parameters of the potential is given by 
\begin{equation} 
\lambda_2(\lambda_2-2\delta)- 4\alpha _4 \, \sqrt{-\epsilon_1}=0, 
\end{equation}
or explicitly, by
\begin{eqnarray}\label{PotConsOrdn1}
&& 64 \alpha_1^2 \alpha_4^8 + 16 \alpha_4^4 (\alpha_3 + 4\alpha_4) (\alpha_3^2 + 
    4 \alpha_3 \alpha_4 + 8 \alpha_4^2 -    4 \alpha_2 \alpha_4^2) \alpha_1  \nonumber \\
    &&\qquad  + (\alpha_3 + 
    4 \alpha_4)^2 (\alpha_3^2 + 2 \alpha_3 \alpha_4 - 
    4 \alpha_2 \alpha_4^2) (\alpha_3^2 + 6 \alpha_3 \alpha_4 +    8 \alpha_4^2 - 4 \alpha_2 \alpha_4^2)
=0.
\end{eqnarray}
Figure~\ref{fig4restrictionsexcited} shows a graph of the allowed values of the parameters $\alpha_2$, $\alpha_3$ and $\alpha_4$, which now are related by \eqref{PotConsOrdn1}, with the restrictions given by the inequalities of \eqref{potV} and \eqref{WavAnsOrd}.
Additionally,  and in a similar way to what was discussed in the previous subsection, Figure~\ref{figpotentialandpsi_first} represents on the left the wave function of the ground state along with its corresponding energy and on the right the specific form of the potential, all for various values of the parameters $(\alpha_2,\alpha_3,\alpha_4)$.

\subsubsection*{\textit{(iii) On the partition function} }
\label{sect:LieOrdn1}
As a natural application of the solutions obtained, we wish to discuss the construction of the canonical partition function, which is the main ingredient in all statistical and thermodynamic studies.
When constructing the partition function for a standard Coulomb problem, unlike what happens for other physical interactions, such as the harmonic oscillator, it is found that the series that defines it is divergent. This divergence has its origin in the nature of wave functions and is actually telling us the series should be truncated and considered only up to a certain finite quantum number.
Such behavior is also present in studies on the simplest generalization, Coulomb potential plus a term with the inverse square, in which the upper limit on the calculation of the partition function is determined from a limiting case of the energy, let's call it the saturation limit (see for example Eqs. (29) and (63) of \cite{Ikot}) assuming a Pekeris type approximation and transforming the problem into an exponential type.
In our case we could anticipate this concept, since the solutions are obtained through the Lie algebraic method and the Bethe ansatz approach, which are intrinsically valid only up to a special quantum number.
The partition function in the usual case is obtained as one in which the constraints between the parameters are already included.
The contribution of the bound state to the partition function at
temperature $T$ with energy \eqref{QESenerExp} is given by 
\begin{equation}\label{partZ}
	 Z=\sum_{n=0}^{\nu} e^{-\epsilon_n/(k_B\, T)} =\sum_{n=0}^{\nu} \exp\left[ \frac{1}{k_B\, T}\frac{\alpha _1^2\,  \alpha _4^2}{(\alpha _3+2 (n+1)\alpha _4)^2}\right] .
\end{equation}
The calculation of the finite sum over $n$ with the maximum value $\nu\in\mathbb{N}$ is performed using the Euler-Maclaurin formula \cite{Euler}  
\begin{equation} \label{EulMac}
\sum_{n=0}^{\nu} f(n)= \int_{0}^{\nu} f(x)dx+\frac{f(\nu)+f(0)}{2}+\sum_{m=1}^{k}\frac{B_{2m}}{(2m)!} \left(f^{(2m-1)}(\nu)-f^{(2m-1)}(0)\right)+R_{\nu,k}  ,
\end{equation}
where $f^{(2m-1)}(x)$ is the derivative of order $2m-1$, $B_{2m}$ are the Bernoulli numbers, $R_{\nu,k}$ is the error term given by an integral involving Bernoulli polynomials, and $k$ is a positive integer. 
The integral appearing in \eqref{EulMac} is 
\begin{equation}
\label{intPart}
I=\frac{\sqrt{\pi } \alpha _1 \left(\text{erfi}\left(\frac{\alpha _1 \alpha _4}{\xi _1 \sqrt{k T}}\right)-\text{erfi}\left(\frac{\alpha _1 \alpha _4}{\sqrt{k T} \left(2 \alpha _4 \nu +\xi _1\right)}\right)\right)}{2 \sqrt{k T}}+\frac{\left(2 \alpha _4 \nu +\xi _1\right) e^{\frac{\alpha _1^2 \alpha _4^2}{k T \left(2 \alpha _4 \nu +\xi _1\right){}^2}}-\xi _1 e^{\frac{\alpha _1^2 \alpha _4^2}{k \xi _1^2 T}}}{2 \alpha _4},
\end{equation}
where $\xi _1=\alpha_3+2\alpha_4$, and $\text{erfi}(z)$ is the imaginary error function $\text{erfi}(z)=\erf(i z)/i$. 

At this point, having obtained the partition function \eqref{partZ}, the thermodynamic properties of the model can be determined, including the free energy $F$, the mean energy $U$, the specific heat $C$ and the entropy $ S$.

However, this is not the right time to delve into these details, which are not the essential objective of our work, and therefore below we will show how the Schr\"odinger equation written in the form \eqref{OrdScheqTrans} can be solved using Heun functions.

\subsection{The ordinary case as a double-confluent Heun equation}\label{sect:HeunOrd}

By changing the independent variable $y= 2 \sqrt{ -\epsilon_n }\,x$, it is found that the differential equation \eqref{OrdScheqTrans} is transformed into a double-confluent Heun equation  \cite{Ronveaux}, that is
\begin{equation}\label{OrdHeun}
	\left\lbrace y^2\frac{d^2}{dy^2}+\left(- y^2+\rho\,y+\eta\right) \frac{d}{dy}-\left(\omega\,y+\lambda_2\right) \right\rbrace\varphi_n (y)=0,
\end{equation}
in which we denote
\begin{equation}
	\rho= 2+\frac{\alpha_3}{\alpha_4},
	\qquad \eta=4 \alpha_4\sqrt{ -\epsilon_n },\qquad 
	\omega= 1+  \frac{\alpha _1}{2 \sqrt{-\epsilon_n}}+\frac{\alpha_3}{2\alpha_4}.
\end{equation}
The regular solutions of \eqref{OrdHeun} at origin, from \cite{Ronveaux,Jaick}, are given by
\begin{equation}\label{HeunFunc}
	\varphi:=\varphi_n (y\,;\rho, \eta,\omega,\lambda_2)=\sum_{n=0}^{\infty} h_n \,x^n,
\end{equation}
in which the coefficients $h_n$ satisfy the relations
\begin{equation*}\label{HeunRecur}
	h_{n+2}=\frac{\left( \lambda_2-n (n+\rho +1)-\rho\right)h_{n+1}+\left(n+\omega\right)h_{n}}{(n+2)\eta}  \qquad \text{and} \qquad h_1=\frac{\lambda_2}{\eta} \,h_0\,,
\end{equation*}
assuming $h_0=1$. 
Consequently, $\varphi$ can admit polynomial solution of degree $m$ if $(m+\omega)$ and $h_{m+1}$ vanish simultaneously. In this way, general solutions to the problem can be obtained in terms of the associated Heun functions.

\section{Solutions of the Coulomb--4 model with a GUP}
\label{sect:BAEGUP}

Having obtained the solutions of the Coulomb--4 model without GUP in the previous section, let us now return to the modified Schr\"odinger equation in a formalism with GUP, that is, to the equations \eqref{GUPeq}--\eqref {VeffDef}. To ensure proper asymptotic behavior of the wave function $\psi_n^{G}(x)$, after inspecting the aforementioned differential equation \eqref{GUPeq}, we propose the following ansatz
\begin{equation}\label{WavAnsGUP}
\psi_n^{G}(x)=x^f\, e^{g(x)}\,\varphi_n^{G}(x) ,\qquad g(x)=-a\,x-\frac{b}{x}-\frac{c}{x^2}-\frac{d}{x^3},
\end{equation}
where the new function $\varphi_n^{G}(x)$ must be a polynomial and the parameters $a>0\,,b\,,c\,,d>0, \, f>0$ are still unknown, but must be such that the original function $\psi_n^{G}(x)$ is square integrable. Substituting \eqref{WavAnsGUP} into \eqref{GUPeq} and solving the Riccati equation that results for $g(x)$, it can be seen that the previous parameters are determined by
\begin{subequations}\label{abcdf}
	\begin{align}
     a&=\sqrt{\gamma_0 } =  \sqrt{  \beta\epsilon_n^2 -\epsilon_n^{G}}>0,\\
     b&= -\frac{ \gamma_7^2-4 \gamma_6 \,\gamma_8}{8\,  (\gamma_8)^{3/2}}=\alpha_2\, \sqrt{ \beta },\\
     c&=  \frac{ \gamma_7 }{ 4  \sqrt{ \gamma_8 }}=   \frac12  \,\alpha_3 \,\sqrt{ \beta },\\
     d&= \frac{\sqrt{\gamma_8 }}{3  }=    \frac13  \alpha_4^2\,   \sqrt{  \beta },\\
     f&= 2+ \frac{ 8 \gamma_5\, \gamma_8^2-4 \gamma_6 \,\gamma_7\, \gamma_8+\gamma_7^3}{16\, (\gamma_8)^{5/2}  } =   2+  \alpha_1 \sqrt{\beta }>0\,.
	\end{align}
\end{subequations}
Then, the differential equation for $\varphi_n^{G}(x)$ simplifies to
\begin{equation}\label{eqGUPphi}
 \left\{P_4(x)\frac{d^2}{dx^2}+Q_4(x)\frac{d}{dx}+W_3(x)\right\}\varphi_n^{G}(x)=0,
\end{equation}
where
\begin{subequations}
	\begin{align}
		P_4(x)=&\ x^4 ,\\[1pt]
		Q_4(x)=&\ 6 d+4 c\, x+2 b\, x^2+2 f\, x^3-2 a \,x^4,\\[1pt]
		W_3(x)= &\ (b^2-6 a d+4 c f-6 c  -\gamma_4 )+  (2 (b f-b-2 a c) -\gamma_3 )x 	\nonumber
		\\
		&\  + (f(f-1)-2 a b -\gamma_2 )x^2-(\gamma_1+2a f) x^3 ,
	\end{align}
\end{subequations}
the parameters $a\,,b\,,c\,,d,\,f$ and $\gamma_i\,$, $i=1,2,3,4$ are given respectively by \eqref{abcdf} and \eqref{gammas}. To solve \eqref{eqGUPphi} we use the general Bethe ansatz method, introduced in Appendix A of Ref.~\cite{AgZh2013}. In this way, we look for polynomial solutions for $\varphi_n^{G}(x)$ of the form
\begin{equation}\label{ansatzphiGUP}
\varphi_n^{G}(x) =\! \left\{
	\begin{array}{cl}
		1, & \quad n=0, \\ 
		\displaystyle\prod_{i=1}^n (x-x_i), & \quad n\in\mathbb{N},
	\end{array}
	\right.  
\end{equation}
where $x_i$ are distinct roots to be determined. Now applying Eqs.~(A.6)--(A.10) of \cite{AgZh2013}, the `general solutions' of \eqref{eqGUPphi} are given by this set of equations:  
\begin{subequations}\label{BAEgeneqs}
	\begin{align}
	& 2a f+2a n +\gamma_1 =0 ,\label{BAEgeneqs1}\\
	& f(f-1)-2 a b-\gamma_2 -2a\sum_{i=1}^{n} x_i+2 f n+(n-1) n=0 ,\label{BAEgeneqs2}\\
	& 2b f-4 a c-2b -\gamma_3 -2a\sum_{i=1}^{n} x_i^2+ 2 (f+n-1)\sum_{i=1}^{n} x_i+2 n\,b=0 ,\label{BAEgeneqs3}\\
	& b^2- 6 a d+4 c f-6 c -\gamma_4 - 2a\sum_{i=1}^{n} x_i^3+2 (f+n-1)\sum_{i=1}^{n} x_i^2+ 2\sum_{i<j}^{n} x_i\,x_j+ 2 b\sum_{i=1}^{n} x_i +4 n\,c=0  ,\label{BAEgeneqs4}
	\end{align}
\end{subequations}	
where $x_i$ are the roots of the  Bethe ansatz equations
\begin{equation}\label{BAroots}
\sum_{j=1,\ j\neq i}^{n} \frac1{x_i-x_j}-\frac{a\, x_i^4 -f\,x_i^3-b\,x_i^2-2c\,x_i-3d}{x_i^4}=0 , \qquad i=1,2,\dots,n. 
\end{equation}
The condition \eqref{BAEgeneqs1} gives the energy relation: for a given $n$, the energy $\epsilon_n^{G}$ is determined by
\small
\begin{equation}\label{nGenEnergGUP}
	\alpha_1  \left[ 1+ 2 \sqrt{ \beta  ( \beta \, \epsilon_n^2 -\epsilon_n^{G} )}-4\,\beta \,\epsilon_n \right]+ 2(n+2)\sqrt{  \beta \, \epsilon_n^2 -\epsilon_n^{G} } =0. 
\end{equation}
\normalsize
Note that the term in square brackets is positive since, according to \eqref{QESenerExp}, $\epsilon_n<0$, again confirming that $\alpha_1$ must be negative for \eqref{nGenEnergGUP} to hold. Through simple manipulations, \eqref{nGenEnergGUP} can be rewritten in closed form like this:
\begin{equation}\label{nGenEnergGUPsimp}
	\epsilon_n^{G}(\epsilon_n;\alpha_1,\beta )=- \frac{\alpha_1^2}4 \left(\frac{ 2\beta  \,\epsilon_n -1}{ \alpha_1\, \sqrt{\beta } +n+2 }\right)^2+ \beta \, \epsilon_n^2 \, .
\end{equation}
Let us keep in mind that the denominator of the fraction, $ \alpha_1\, \sqrt{\beta } +n+2$, is always positive because  $f>0 $, as established in \eqref{abcdf}, and remember that $\epsilon_n$ is given by \eqref{QESenerExp}. On the other hand, the equations \eqref{BAEgeneqs2}--\eqref{BAEgeneqs4} produce severe restrictions on the potential parameters, explicitly:
\begin{subequations}\label{alphas,nGen} 
	\begin{align}
		\alpha_2&= \frac{ 2  \sqrt{  \beta  \,\epsilon_n^2-\epsilon_n^{G}} 
	\sum\limits^{n}_{i=1} x_i-  \alpha_1 \sqrt{ \beta }  (2 n+3)-(n+1) (n+2) 
	 }{2\beta\,\epsilon_n - 2 \sqrt{\beta   ( \beta  \,\epsilon_n^2 -\epsilon_n^{G})}-1}   ,\\[1ex]  
	\alpha_3&=  2   \, \frac{ \sqrt{  \beta  \,\epsilon_n^2-\epsilon_n^{G}}
	\sum\limits^{n}_{i=1} x_i^2-  \sqrt{ \beta } \Bigl( (n+1)\alpha_2 +\alpha_1 \sum\limits^{n}_{i=1} x_i \Bigr) -(n+1)
	 \sum\limits^{n}_{i=1} x_i 
	}{2\beta\,\epsilon_n - 2 \sqrt{\beta   ( \beta  \,\epsilon_n^2 -\epsilon_n^{G})}-1}     , \\[1ex]   
	\alpha_4^2&= \frac{ 2 \sqrt{ \beta  \,\epsilon_n^2-\epsilon_n^{G} }
	\sum\limits^{n}_{i=1} x_i^3- \sqrt{  \beta }  \Bigl( (2n+1) \alpha_3 +2 \alpha_2 \sum\limits^{n}_{i=1} x_i+2 \alpha_1 \sum\limits^{n}_{i=1} x_i^2 \Bigr) - 2(n+1)\sum\limits^{n}_{i=1} x_i^2-2\sum\limits^{n}_{i<j} x_i\,x_j }{2\beta\,\epsilon_n - 2 \sqrt{\beta   ( \beta  \,\epsilon_n^2 -\epsilon_n^{G})}-1}  \,.
\end{align}
\end{subequations}	
The denominators of the fractions are again non-zero since $\epsilon_n<0$ by \eqref{QESenerExp}. The parameters $x_i$ are determined by the Bethe ansatz equations
\begin{equation}\label{BArootsExp}
			\sum_{j=1,\, j\neq i}^{n} \frac1{x_i-x_j}-\frac{ x_i^4 \sqrt{ \beta \,\epsilon_n^2-\epsilon_n^{G} }- \sqrt{  \beta } \left(\alpha_4^2+\alpha_1\, x_i^3+\alpha_2\, x_i^2+\alpha_3\, x_i\right)-2 x_i^3  }{x_i^4}=0 , \quad i=1,2,\dots,n. 
\end{equation}
In summary, for a given $n$, the `general solutions' of the Schr\"odinger equation modified with a GUP are given by the equations \eqref{WavAnsGUP} together with \eqref{ansatzphiGUP} and \eqref{nGenEnergGUPsimp}--\eqref{BArootsExp}. As in the ordinary case (without a GUP),  we will now look for explicit solutions for the ground state and the first excited state.

\subsubsection*{\textit{(i)} Ground state}
\label{sect:GUPn0}
For $n=0$, from \eqref{nGenEnergGUPsimp} the energy of ground state, $\epsilon_0^{G}$, is determined in closed form:
\begin{equation}\label{n0EnergGUPsimp}
	\epsilon_0^{G}(\epsilon_0;\alpha_1,\beta )=- \frac{\alpha_1^2}4 \left(\frac{ 2\beta  \,\epsilon_0 -1}{ \alpha_1\, \sqrt{\beta } +2 }\right)^2+ \beta \, \epsilon_0^2 \, ,
\end{equation}
where the ground state energy of the ordinary case, $\epsilon_0$, was already obtained in \eqref{QESenerExpn0}. From \eqref{WavAnsGUP} together with \eqref{abcdf} and \eqref{ansatzphiGUP}, the explicit form of the associated wave function is given by
\begin{equation}\label{n0WaveGUP}
	\psi_0^{G}(x)=C_0\ x^{2+ \alpha_1\sqrt{ \beta } }\ \exp\left[ 
	 -  x\, \sqrt{\beta  \, \epsilon_0^2 -\epsilon_0^{G}} - \frac{\sqrt{\beta}}{2} \left( \frac{2 \alpha_4^2}{3 x^3}+\frac{\alpha_3}{x^2}+\frac{2 \alpha_2}{x} \right) \right] ,
\end{equation}
where $C_0$ is the normalization constant and, due to \eqref{alphas,nGen}, the parameters of the potential satisfy the constraints
\begin{subequations}\label{alphas,n0}
\begin{align}
	\alpha_2&= \frac{  2+ 3\sqrt{ \beta } \,\alpha_1}{2 \sqrt{\beta  \left( \beta  \,\epsilon_0^2 -\epsilon_0^{G}\right)}-2 \beta\,\epsilon_0+1} , \label{tururu}\\[1ex]
	\alpha_3&= \frac{2 \sqrt{ \beta } }{2 \sqrt{\beta  \left( \beta  \,\epsilon_0^2 -\epsilon_0^{G}\right)}-2 \beta\,\epsilon_0+1}\ \alpha_2, \\[1ex]
	\alpha_4^2&= \frac{\sqrt{ \beta } }{2 \sqrt{\beta  \left( \beta  \,\epsilon_0^2 -\epsilon_0^{G}\right)}-2 \beta\,\epsilon_0+1}\ \alpha_3\, .
\end{align}
\end{subequations}
Note that the Bethe ansatz equations \eqref{BAroots}  play no role in the ground state solution given the fact that $ \varphi_0^{G}(x) \equiv 1$, cf. \eqref{ansatzphiGUP}. 
It can also be seen that the above restrictions imply that $\alpha_2$ and $\alpha_3$ are positive in the case of GUP, taking into account that $\epsilon_0<0$.
As a consequence, from the  equation \eqref{tururu}, we find that $-\frac{2 }{3 \sqrt{  \beta }  } <\alpha_1<0 $.

\subsubsection*{\textit{(ii)} First excited state}
\label{sect:GUPn1}

For $n=1$, from \eqref{nGenEnergGUPsimp}, the ground state energy, $\epsilon_1^{G}$, is given by the closed form expression
\begin{equation}\label{n1EnergGUPsimp}
	\epsilon_1^{G}(\epsilon_1;\alpha_1,\beta )=- \frac{\alpha_1^2}4 \left(\frac{ 2\beta  \,\epsilon_1 -1}{ \alpha_1\, \sqrt{\beta } + 3 }\right)^2+ \beta \, \epsilon_1^2 \, .
\end{equation}
provided $\alpha_1<0$. Recall again that $\epsilon_1$ was already obtained in \eqref{QESenerExpn1}. 
In the present case it is assumed that the differential equation \eqref{eqGUPphi} has solutions in the form of \eqref{ansatzphiGUP}, that is, $\varphi_1^{G}(x)=x-x_1\,$, and consequently, from \eqref{WavAnsGUP}--\eqref{abcdf}, the wave function is explicitly given by  
 \begin{equation}\label{n1WaveGUP}
 	\psi_1^{G}(x)=C_1\ (x-x_1)\ x^{2+ \alpha_1\sqrt{ \beta } }\ \exp\left[ 
	 -  x\, \sqrt{\beta  \, \epsilon_1^2 -\epsilon_1^{G}} - \frac{\sqrt{\beta}}{2} \left( \frac{2 \alpha_4^2}{3 x^3}+\frac{\alpha_3}{x^2}+\frac{2 \alpha_2}{x} \right) \right] ,
 \end{equation}
 where $C_1$ is the normalization constant.
In addition, from \eqref{alphas,nGen}, we have the following constraints on the potential parameters
\begin{subequations}\label{alphas,n1}
	\begin{align}
	\alpha_2&= \frac{ 2\, x_1\, \sqrt{ \beta\,\epsilon_1^2 -\epsilon_1^{G} }-5\,\alpha_1\, \sqrt{\beta } -6 }{-2 \sqrt{\beta \left( \beta  \,\epsilon_1^2 -\epsilon_1^{G}\right)}+2 \beta\,\epsilon_1 -1} \,,\\[7pt]
	\alpha_3&=  \frac{ 2\, x_1^2\, \sqrt{ \beta\,\epsilon_1^2 -\epsilon_1^{G} } -2 \sqrt{\beta } (2\alpha_2+\alpha_1\,x_1)-4\,x_1}{-2 \sqrt{\beta \left( \beta  \,\epsilon_1^2 -\epsilon_1^{G}\right)}+2 \beta\,\epsilon_1 -1} \,, \\[7pt]
	\alpha_4^2&=  \frac{ 2\, x_1^3\, \sqrt{ \beta\,\epsilon_1^2 -\epsilon_1^{G} }  - \sqrt{\beta }\, \big(3\alpha_3+2(\alpha_2+\alpha_1\,x_1)x_1\big)- 4\,x_1^2}{-2 \sqrt{\beta \left( \beta  \,\epsilon_1^2 -\epsilon_1^{G}\right)}+2 \beta\,\epsilon_1 -1} \, .
		\end{align}
\end{subequations}
The unknown parameter $x_1$, which appears in \eqref{n1WaveGUP} and \eqref{alphas,n1}, is determined from the Bethe ansatz equation \eqref{BAroots}, which turns out to be
\begin{equation}\label{BAroots,n1}
\sqrt{ \beta\, \epsilon_1^2 -\epsilon_1^{G} }\, x_1^4 
-  \left( 2+\alpha_1 \sqrt{ \beta } \right) \, x_1^3 - \sqrt{\beta } \left(\alpha_2  \,x_1^2+  \alpha_3  \,x_1+  \alpha_4^2 \right) =0\,.
\end{equation}

\section{Conclusions}
\label{sect:conc}

Based on recent papers proposing a correction of the Coulomb potential by including inverse quadratic, cubic and  quartic terms, we consider this modified potential in the framework of the modified Schr\"odinger equation with a generalized uncertainty principle. 
The resulting equation appears as a generalization of the Heun equation, more precisely, the biconfluent Heun equation that has the form $y''(x)+\left(A_0+\frac{A_{-1}}{x}+ \frac{A_{-2}}{ x^2}+\frac{A_{-3}}{x^3}+\frac{A_{-4}}{x^4}\right)y(x )=0$ \cite{Ronveaux}. 
To the best of our knowledge, the general form we consider could not be solved by other analytical techniques, including the common Lie algebraic approach, which requires the form $P_4(x)\Phi''(x)+Q_3(x)\Phi'(x)+W_2 ( x ) \Phi(x)=0$ \cite{Turbiner,Turbiner88}. 
Therefore, we attempted a Bethe analysis of the equation and after some calculations we found the general solution to the modified problem due to the GUP and then both the ground state and the first excited state are explicitly determined.
It is worth mentioning that, despite its merits, this approach is very burdensome for higher states, since the determination of the parameters involved becomes more complicated, due to the restrictions that must be imposed on the parameters in order to provide solutions. However, the truth is that this last limitation exists in most approaches, including supersymmetric quantum mechanics, power series solutions, integral transforms, etc.

Let us now review our solutions to realize that we have already given the solutions to some important physical problems. The first application to consider is the simple but rich toy model of the mini-universe proposed by Roshan \cite{Roshan} to study the expansion (or contraction) of a homogeneous and isotropic universe. 
Equation (27) from Roshan's paper, which is the governing quantum equation, can be converted to that of our problem by a transformation $\phi(x)=R(x)\psi(x)$. 
Now, having calculated the solutions of the ordinary cases of the so-called generalized Coulomb--4 potential, all the results there, including the mass and the Bohr radius, can now be derived simply with a quite realistic potential, which includes both post-Newtonian relativistic and quantum corrections.  

A second application of our work is to the so-called Vaz quantum gravitational model. In a very recent paper \cite{Corda}, Corda reviews the firewall paradox and Vaz's quantum gravitational model \cite{Corda}.
The Vaz quantum mechanical model was originally formulated to study the spherical collapse of inhomogeneous dust and is AdS of dimension $d=n+2$.
The author considers a two-body Hamiltonian and, having separated the problem into a center-of-mass framework, derives a Schrödinger-type equation with an inverse term plus a gravitational potential, which is also Coulomb-type.
The potential considered there is the ordinary gravitational potential, the energy of which determines the mass and energy spectra of the black hole and the density of the Vaz shell.
Based on the results obtained here for the ordinary case of the one-dimensional Schrödinger equation, one can simply revise that model with the generalized gravitational potential with a simple parameter change in the calculations.

A third similarity to the structure analyzed in the present work arises with a semi-relativistic approximation of a quantum field theoretical two-body Bethe-Salpeter equation, which is often called the spinless Salpeter equation, and finds applications in a variety of physical fields \cite{Lucha,Adamo}.

A fourth category of problems that is related to our work is that of wave equations in curved space or wave equations in the space-time of cosmic strings, such as, for example, the three-dimensional Dirac equation in a rather general curved spacetime \cite{Jahangiri,Medeiros}.
It should be noted that in the similar cases mentioned, there is not as much difficulty as here in treating the associated ordinary energy levels as a direct parameter in the calculation.

In conclusion, we think that the present work provides a basis for addressing some important problems in other related fields and that it could be useful for further studies. In particular, a generalization of the GUP with this interaction to a three-dimensional case and a two-body formulation will be a suitable framework for future theoretical investigations or probably for designing experimental setups. The latter could be particularly attractive in relation to Planck-scale physics and quantum optics \cite{Pikovski}, since a direct and realistic quantum term is now available, which needs further consideration as to its exact value.

\section*{Acknowledgments}

The research of L.M.N. and S.Z. was supported by 
the European Union.-Next Generation UE/MICIU/Plan de Recuperacion, Transformacion y Resiliencia/Junta de Castilla y Leon,
RED2022-134301-T financed by MICIU/AEI/10.13039/501100011033, 
and PID2020-113406GB-I00 financed by MICIU/AEI/10.13039/501100011033.
The work of M.B. was supported by the Czech Science Foundation within the project 22-18739S.

\end{document}